\documentclass[aps,prl,preprint,nopacs,superscriptaddress]{revtex4}

\usepackage{graphicx}
\usepackage{verbatim}
\usepackage{mathrsfs}
\usepackage{amsmath,amsfonts,amssymb}
\usepackage{graphicx}

\def\3{2.8in}    
\def\2{2.5in}
\def\4{3.0in}

\def \beq {\begin{equation}}
\def \eeq {\end{equation}}
\begin{document}

\title{Observation of Fermi Arc Surface States in a Topological Metal: A New Type of 2D Electron Gas}

\author{Su-Yang Xu}
\affiliation {Joseph Henry Laboratory, Department of Physics, Princeton University, Princeton, New Jersey 08544, USA}
\author{Chang Liu}
\affiliation {Joseph Henry Laboratory, Department of Physics, Princeton University, Princeton, New Jersey 08544, USA}
\author{Satya K. Kushwaha}
\affiliation {Department of Chemistry, Princeton University, Princeton, New Jersey 08544, USA}
\author{Raman Sankar} \affiliation{Center for Condensed Matter Sciences, National Taiwan University, Taipei 10617, Taiwan}

\author{Jason W. Krizan}
\affiliation {Department of Chemistry, Princeton University, Princeton, New Jersey 08544, USA}

\author{Ilya Belopolski}
\affiliation {Joseph Henry Laboratory, Department of Physics, Princeton University, Princeton, New Jersey 08544, USA}

\author{Madhab Neupane}
\affiliation {Joseph Henry Laboratory, Department of Physics, Princeton University, Princeton, New Jersey 08544, USA}
\author{Guang Bian}
\affiliation {Joseph Henry Laboratory, Department of Physics, Princeton University, Princeton, New Jersey 08544, USA}

\author{Nasser Alidoust}
\affiliation {Joseph Henry Laboratory, Department of Physics, Princeton University, Princeton, New Jersey 08544, USA}
%
%
%
%

\author{Tay-Rong Chang}
\affiliation{Department of Physics, National Tsing Hua University, Hsinchu 30013, Taiwan}

\author{Horng-Tay Jeng}
\affiliation{Department of Physics, National Tsing Hua University, Hsinchu 30013, Taiwan}
\affiliation{Institute of Physics, Academia Sinica, Taipei 11529, Taiwan}

\author{Cheng-Yi Huang}
\affiliation{Department of Physics, National Sun Yat-Sen University, Kaohsiung 804, Taiwan}

\author{Wei-Feng Tsai}
\affiliation{Department of Physics, National Sun Yat-Sen University, Kaohsiung 804, Taiwan}

\author{Hsin Lin}
\affiliation {Graphene Research Centre and Department of Physics, National University of Singapore 11754, Singapore}

\author{Fangcheng Chou} \affiliation{Center for Condensed Matter Sciences, National Taiwan University, Taipei 10617, Taiwan}

\author{Pavel P. Shibayev}
\affiliation {Joseph Henry Laboratory, Department of Physics, Princeton University, Princeton, New Jersey 08544, USA}

\author{Robert J. Cava}
\affiliation {Department of Chemistry, Princeton University, Princeton, New Jersey 08544, USA}

\author{M. Zahid Hasan}
\affiliation {Joseph Henry Laboratory, Department of Physics, Princeton University, Princeton, New Jersey 08544, USA}

\pacs{}

\date{\today}

\begin{abstract}
\textbf{In a topological insulator, it is the electrons on the surface or edge that carry the signature of topology. Recently, a novel topological state has been proposed in metals or semimetals whose band-structure is similar to that of a three-dimensional analog of graphene. However, to this date the signature of its topology remains an open question. We report the experimental discovery of a pair of polarized Fermi arc surface state modes in the form of a new type of two-dimensional polarized electron gas on the surfaces of Dirac semimetals. These Fermi arc surface states (FASS) are observed to connect across an even number of bulk band nodes and found to have their spin uniquely locked to their momentum. We show that these states are distinctly different from the topological surface states (TSS) seen in all topological insulators. Our observed exotic two-dimensional states not only uncover the novel topology of Dirac metals (such as sodium tribismuth Na$_3$Bi) but also opens new research frontiers for the utilization of Fermi arc electron gases for a wide range of fundamental physics and spintronic studies envisioned in recent theories.}
\end{abstract}
\maketitle

\section{Introduction}

Realization of topological states of matter beyond topological insulators has become an important goal in condensed matter and materials physics \cite{Ashvin_Review,Haldane, Nagaosa, 3D_Dirac, Dirac_3D, Dirac_semi, Dai, Ashvin, Ashvin2, Hosour, Ando, Ojanen, RMP, Zhang_RMP, Kim, Geim,Hasan_Na3Bi}. In topological insulators Bi$_{1-x}$Sb$_x$ and Bi$_2$Se$_3$ or topological crystalline insulators such as the Pb$_{1-x}$Sn$_x$Te(Se), the bulk has a full insulating energy gap, whereas the surface possesses an odd or even number of spin polarized surface or edge states \cite{RMP, Zhang_RMP}. These are symmetry-protected topological states \cite{Wen}. It is the 2D surface state or 1D edge state properties that carry the signature of topology, which is the key focus of experiments on all topological states of matter. Very recently, the possibility of realizing new topological states beyond insulators, such as metals or semimetals, has attracted much attention \cite{Ashvin_Review,Haldane, Nagaosa, 3D_Dirac, Dirac_3D, Dirac_semi, Dai, Ashvin, Ashvin2, Hosour, Ando, Ojanen}. Semimetals are materials, whose bulk conduction and valence bands have small but finite overlap. Thus there does not exist a full band gap irrespective of the choice of the chemical potential. Since the definition of bulk topological number for topological insulators strictly requires a full bulk energy gap \cite{RMP, Zhang_RMP}, any topological states that might exist in a semimetal should be fundamentally distinct from the topological states studied in insulating materials. Therefore, the identification of topological boundary modes in semimetals opens a new era in topological physics \cite{Ashvin_Review,Haldane, Nagaosa}, which is analogous to the first identification of topological surface states in topological insulators. Up until now, theory has proposed two kinds of topological semimetals, the topological Dirac and Weyl semimetals \cite{Nagaosa, 3D_Dirac, Dirac_3D, Dirac_semi, Dai, Ashvin}. Their low energy bulk excitations are described by the Dirac and Weyl equations, respectively. And for both types, the bulk is predicted to show linear band touching points between the conduction and valence bands at multiple discrete points in the momentum space or Brillouin zone \cite{Nagaosa, 3D_Dirac, Dirac_3D, Dirac_semi, Dai, Ashvin}. However, the topological distinction between semimetals and insulators is defined through the distinction between the surface states that they actualize. For a topological semimetal, the decisive signature for nontrivial topology is a pair of Fermi arc states that link an even number of bulk nodes. Therefore, the full Fermi surface of a topological semimetal consists of both the exotic Fermi arc surface states (FASS) and the bulk nodes, which is fundamentally different from that of a TI, whose full Fermi surface is a circle-like contour of surface states (TSS) only. The double Fermi arc surface states that uniquely exist in nontrivial semimetals are disjoint on the surface, where they can only connect to each other through the bulk bands in the form of a closed loop. This 2D surface state, which serves as the topological signature of nontrivial metals or semimetals, has never been observed in experiments since all the experiments so far have focused on the basic bulk band structures revealing just 3D analogs of graphene without probing any topological content. The existence of Fermi arc surface states connecting even number of linear bulk band touching points defines the topology and topological content of the nontrivial semimetal phases and requires finding a suitable surface termination as well. These exotic properties can further give rise to a wide range of new phenomena in transport and magnetization measurements, which can only be observed once the correct surface termination is also identified. For example, the linear dispersive bulk band touching is predicted to show a chiral anomaly \cite{Ando}, while the realization of multiple bulk Dirac or Weyl nodes can lead to a quantum Lifshitz transition in momentum space - a form of electronic singularity \cite{Kim, Geim, Hasan_Na3Bi} leading to correlated electronic states. For the Fermi arc surface states, theory has predicted spin polarization \cite{Ojanen}, Friedel oscillations \cite{Hosour}, and novel quantum oscillation \cite{Ashvin} in transport and spectroscopic experiments, where magnetic field is to be applied along certain surface normals. These new phenomena are not possible to realize in the much studied 3D topological insulators.


Experimentally, a number of compounds have been identified to be 3D Dirac semimetals such as BiTl(S$_{0.5}$Se$_{0.5}$)$_2$, (Bi$_{0.94}$In$_{0.06}$)$_2$Se$_3$, BiNa$_3$ (or Na$_3$Bi), and Cd$_3$As$_2$ \cite{Suyang, Oh, Chen_Na3Bi, CdAs_Hasan, CdAs_Cava}. All these compounds exhibit nearly linear bulk band crossing near the Fermi level and their bulk band structure is now well-known \cite{Suyang, Oh, Chen_Na3Bi, CdAs_Hasan, CdAs_Cava}. However, based on the vast existing experimental data on 3D Dirac semimetals, one cannot conclude their topological nature because Fermi arc surface states have never been observed on the crystallographic surfaces that have been studied despite a lot of experimental activities worldwide. In this paper, we identify a suitable and new surface termination that can reveal the topology of Dirac semimetals by combining both angle and spin-resolved measurements. We experimentally uncover the nontrivial topological nature in semimetal bismuth trisodium (BiA$_3$, A is an alkali metal, A = Na in our case), through a careful study of its (100) surface (since the spin-orbit interaction originates from the Bi-atoms, which is relevant for the topology, we write this compound as BiNa$_3$). Using high-resolution angle-resolved photoemission spectroscopy (ARPES) we obtain critical experimental data revealing that this compound hosts a pair of linear band touching points in its bulk and double Fermi arc surface states (FASS) on its (100) surface. The double surface Fermi arcs are observed to connect the two bulk linear band touching points and exhibit spin polarization and spin momentum locking at the momentum space locations away from the bulk nodes. We further present a systematic study revealing the properties of these new topological surface states in semimetals. These results identify the topological semimetal and our observation of double Fermi arc surface states and their unique spin momentum locking demonstrate a topological number of $\nu_{\textrm{2D}}=1$ for the $k_{[001]}=0$ 2D slice that defines the topology of this metal. Our systematic results not only lays the foundation for opening the door for studying many fundamentally new physics in nontrivial metals (not insulators), but also offers an entirely new type of 2D electron gas with potential for spin manipulation \cite{Ashvin_Review,Haldane, Nagaosa, 3D_Dirac, Dirac_3D, Dirac_semi, Dai, Ashvin, Ashvin2, Hosour, Ando, Ojanen, RMP, Zhang_RMP,Kim, Geim, Hasan_Na3Bi}.

\section{Results}

\textbf{The topological semimetal phase with nonzero invariant}

Bismuth trisodium (BiNa$_3$) is a semimetal that crystalizes in the hexagonal $P6_3/mmc$ crystal structure with $a=5.448$ $\textrm{\AA}$ and $c=9.655$ $\textrm{\AA}$ \cite{Na3Bi_crystal}. First principles bulk band calculations \cite{Dirac_semi} show that its lowest bulk conduction and valence bands are composed of Bi $6p_{x,y,z}$ and Na $3s$ orbitals. These two bands possess a bulk band inversion of about $\sim0.3$ eV at the bulk BZ center $\Gamma$ \cite{Dirac_semi}. The strong spin-orbit coupling in the system can open up energy gaps between the inverted bulk bands, but due to the protection of an additional three-fold rotational symmetry along the [001] crystalline direction, two bulk Dirac band touchings (Dirac nodes) are predicted to be preserved even after considering spin-orbit coupling, as schematically shown by the blue crosses in Fig.~\ref{Arc}\textbf{F}. The two bulk Dirac nodes locate along the $A$-$\Gamma$-$A$ [001] direction. Therefore, at the (001) surface, the two bulk Dirac nodes project onto the same point in the (001) surface BZ (Fig.~\ref{Arc}\textbf{F}), making it difficult  to separate, isolate, and systematically study them via spectroscopic methods. More importantly, Fermi arc surface states are fundamentally not possible because a Fermi arc starts from one bulk Dirac node and ends on another (therefore requires multiple bulk Dirac nodes that are separated within the surface BZ). On the other hand, at the (100) surface, the two bulk Dirac nodes are separated on the opposite sides of the (100) surface BZ center $\tilde{\Gamma}$ (Fig.~\ref{Arc}\textbf{F}). Consequently, the Fermi arc surface states that connect the bulk Dirac nodes are found in the (100) surface electronic structure calculation (Fig.~\ref{Arc}\textbf{H}). The surface states are double Fermi arcs because they exist everywhere on the Fermi surface contour (Fig.~\ref{Arc}\textbf{D}) except at the two bulk Dirac points. In other words, if one travels around the Fermi surface contour, the electronic states are localized at the surface everywhere, except at the locations of the two bulk Dirac points where the states disperse linearly in all three dimensions. We note that the existence of double Fermi arc surface states in bismuth trisodium and cadmium arsenide has been theoretically predicted by a number of theory works \cite{Nagaosa, Dirac_semi, Dai, Ashvin2}, which strongly motivated our work to experimentally search for Fermi arc surface states in Dirac metals.

Due to the absence of a full bulk energy gap in semimetal materials, it is not possible to associate a topological invariant with the entire 3D Brillouin zone for a topological semimetal system. However, because the bulk gap only closes at the Weyl/Dirac points, most 2D slices of the 3D Brillouin zone are fully gapped. In order to characterize the topological property of the topological semimetal phases, one can define 2D topological number for the 2D $k$-slices that have a full bulk energy gap \cite{Nagaosa, 3D_Dirac}. For the topological Weyl semimetal phase, the 2D topological number for the 2D $k$-slice is the Chern number, which does not need any additional symmetry protection \cite{3D_Dirac}. For the topological Dirac semimetal phase, the 2D topological number can be the 2D Z$_2$ number $\nu_{\textrm{2D}}$ or the mirror Chern number $n_M$, which is protected by time-reversal or crystalline mirror symmetry \cite{Nagaosa}. Specifically, in the case of bismuth trisodium, it has been theoretically shown that there exists a 2D Z$_2$ topological number $\nu_{\textrm{2D}}=1$ in the $k_{[001]}=0$ plane \cite{Nagaosa}, which manifests the topological nontriviality of this system. In order to experimentally probe the topological property of bismuth trisodium, we systematically study its electronic structure and spin polarization at the (100) side-surface. We note that our previous ARPES results (unpublished) \cite{Hasan_Na3Bi} on the (100) side-surface were obtained from a different batch of relatively $p$-type samples, where the Fermi arc surface states are above the Fermi level and therefore were not observed. Here we report the first observation of Fermi arc surface states.

\bigskip
\textbf{Spin-polarized double Fermi arc surface states}

Fig.~\ref{Arc}\textbf{A} shows the ARPES measured Fermi surface of our bismuth trisodium sample at its native Fermi level. Remarkably, the measured Fermi surface is found to consist of two Fermi ``points'' along the $k_y(100)$ direction and two arcs that connect the two Fermi points. The measured Fermi surface topology (Fig.~\ref{Arc}\textbf{A}) is in agreement with the theoretical prediction \cite{Dirac_semi}, where two surface Fermi arcs connect the two bulk nodal touchings. In order to confirm that the observed Fermi surface is indeed the double Fermi arcs connecting two bulk nodal points, we study the evolution of constant energy contour as a function of binding energy $E_{\textrm{B}}$. As shown in Fig.~\ref{Arc}\textbf{B}, the energy contour area of the two bulk nodal touching points is found to enlarge into contours as $E_{\textrm{B}}$ is increased (hole-like behavior), whereas the two surface Fermi arcs shrink while increasing $E_{\textrm{B}}$ (electron-like behavior). The evolution of the constant energy contour as a function of binding energy $E_{\textrm{B}}$ (electron or hole behavior for different bands) is also consistent with the theoretical expectation (see Fig.~\ref{Arc}\textbf{D}). To further confirm, we study the energy dispersion for important momentum space cut directions (Fig.~\ref{Arc}\textbf{B}). As shown in Fig.~\ref{Arc}\textbf{C}, surface states with a surface Dirac crossing are clearly observed near the Fermi level in Cut $\beta$, consistent with the theoretical calculation. The bulk valence band for Cut $\beta$ is found to be away from the Fermi level. On the other hand, for Cuts $\alpha$ and $\gamma$, no surface states are observed but the bulk linear band is seen to cross the Fermi level, also consistent with the calculation (see Fig.~\ref{Arc}\textbf{E}). The contrasting behavior that for Cut $\beta$ only surface states cross the Fermi level whereas for Cuts $\alpha$ and $\gamma$ only the bulk linear bands cross the Fermi level serves as a piece of critical evidence for the existence of double Fermi arc surface states. We note that the two bulk nodal touchings observed in Fig.~\ref{Arc}\textbf{A} still expand in a finite area in momentum space, rather than being ideal ``points''. This is because (1) electronic states in real samples have finite quasi-particle life time and mean free path, and (2) it is also possible that the chemical potential of our bismuth trisodium sample is still slightly below the energy of the bulk nodes.

In order to further confirm the 2D surface nature for the double Fermi arc surface states and the 3D bulk nature for the two bulk Dirac bands, we present Fermi surface maps at different incident photon energies. Upon varying the photon energy, one can study the electronic dispersion along the out-of-plane $k_z$ direction. Due to the different 2D $vs$ 3D nature of the Fermi arc surface states and the bulk Dirac bands, the Fermi arc surface states are expected to be independent of $k_z$ (photon energy), whereas the bulk bands are expected to show strong $k_z$ dependence. As shown in Fig.~\ref{hv_spin}\textbf{A}, the double Fermi arcs are clearly observed irrespective of the choice of the photon energy values, which confirms their 2D surface nature. On the other hand, the two bulk nodal points are found to be quite pronounced at photon energies of 58 eV and 55 eV, but they become nearly unobservable as the photon energy is changed to 40 eV (e.g. the area highlighted by the red dotted circles in Fig.~\ref{hv_spin}\textbf{A}). This demonstrates the 3D bulk nature of the two bulk bands. To quantitatively evaluate how much the bulk band becomes weaker as $h\nu$ is changed from 58 eV to 40 eV (its $k_z$ dependence), we denote the middle point of the left Fermi arc as ``S'' and the bottom bulk nodal touching point as ``B'' (Fig.~\ref{hv_spin}\textbf{B}). Fig.~\ref{hv_spin}\textbf{C} shows the relative ARPES intensity between ``S'' and ``B''. The strong dependence of $\frac{I(\textrm{B})}{I(\textrm{S})}$ upon varying photon energy shows that the bulk Dirac band becomes much weaker relative to the surface states as $h\nu$ is changed from 58 eV to 40 eV, due to the reason that the $k_z$ value is moved away from the bulk nodal (Dirac) point. Furthermore, Fig.~\ref{hv_spin}\textbf{D} shows the dispersion along Cut $\beta$, where the surface states with a clear Dirac crossing are observed at different photon energies. This confirms the 2D nature of the Fermi arc surface states along Cut $\beta$. In Fig.~\ref{hv_spin}\textbf{E}, we present photon energy dependent dispersion of the bulk nodal band (cut indicated by the white dotted line in Fig.~\ref{hv_spin}\textbf{A}). It can be seen that the bulk Dirac band only crosses the Fermi level at a photon energy around 58 eV and disperses strongly upon varying the photon energy ($k_z$) value, which confirms the 3D bulk nature of the bulk Dirac band. These systematic photon energy ($k_z$) dependent data, shown in Fig.~\ref{hv_spin}, clearly demonstrate that the electronic states at the Fermi level are localized at the surface (2D surface state nature) everywhere around the Fermi surface contour, except at the locations of the two bulk nodal touching points, where the states disperse strongly in all three dimensions (3D bulk band nature). These data sets provide further support for the observation of double Fermi arc surface states in bismuth trisodium. We study the surface spin polarization along the Cut $\beta$. The white lines in Fig.~\ref{Gapclosing}\textbf{A} define the two momenta chosen for spin-resolved studies (namely S-Cut1 and S-Cut2). Spin-resolved measurements are performed at these two fixed momenta as a function of binding energy. Figs.~\ref{Gapclosing}\textbf{B,C} show the in-plane spin-resolved intensity and net spin polarization. The magnitude of spin polarization reaches about $30\%$ near the Fermi level. And the direction of spin polarization is reversed as one goes from S-Cut1 to S-Cut2, which shows the spin-momentum locking property and the singly degenerate nature of the Fermi arc surface states along the Cut $\beta$ direction.

\bigskip
\textbf{Topological invariant for the semimetal}

We use the obtained electronic and spin data for the (100) Fermi arc surface states to gain insights to the predicted 2D topological number ($\nu_{\textrm{2D}}=1$) in bismuth trisodium. In Figs.~\ref{Gapclosing}\textbf{D-F}, we present a series of ARPES dispersion maps. As schematically shown in Fig.~\ref{Gapclosing}\textbf{D}, these maps, which are perpendicular to the $k_{[001]}$ axis, intersect with the axis at different $k_{[001]}$ values. As one goes from Slice1 to Slice7 along the $k_{[001]}$ axis, the ARPES measured and schematic energy dispersion for different sides are shown in Figs.~\ref{Gapclosing}\textbf{E,F}. Due to the existence of the two bulk nodes (Slices 2 and 6), the bulk band gap closes and reopens as one goes across each bulk node. Therefore, it is interesting to note that the $k_{[001]}$ axis serves as an effective axis for a bulk mass parameter in a 2D system. It is further interesting to note that the bulk band gap closings at Slices 2 and 6 correlated with the absence/appearance of the surface states at the Fermi level. As clearly seen in Figs.~\ref{Gapclosing}\textbf{E,F}, there are no surface states for Slice1. As one moves from Slice1 to Slice3, the bulk band gap closes and reopens, and surface states appear at the Fermi level of bismuth trisodium. Similarly, as one further goes from Slice3 to Slice7, the bulk band gap closes and reopens, and the surface states do not exists anymore for the data in Slice7. We note that the gap closing and reopening property that we observed here does not exist in the Fermi surface of any known topological insulator system such as Bi$_2$Se$_3$, which again highlights that the observed Fermi arc surface states (FASS) are different from the topological surface states (TSS) in a topological insulator.

We focus on Slices 3-5, which are in-between the two bulk nodes. Interestingly, as seen in Figs.~\ref{Gapclosing}\textbf{E,F}, although surface states exist at the native Fermi level for any Slices that are in-between the two bulk nodes, they are in general gapped (e.g. Slice3 in Figs.~\ref{Gapclosing}\textbf{E,F}). The gapped nature of the surface states means that, for these Slices ($k_{[001]}\neq0$), there does not exist a nontrivial 2D topological number that can protect a doubly degenerate Dirac crossing for the surface states. However, for Slice4, which corresponds to $k_{[001]}=0$, the surface states are observed to be gapless with a surface Dirac point. Furthermore, our spin-resolved data in Figs.~\ref{Gapclosing}\textbf{A-C} clearly show that the two branches of the surface Dirac crossing carry opposite spin polarization. The gapless surface Dirac crossing and the spin-momentum locking for Slice4 ($k_{[001]}=0$), collectively, provide experimental evidence that the $k_{[001]}=0$ plane has a nontrivial 2D topological number, which is consistent with the $\nu_{\textrm{2D}}=1$ for the $k_{[001]}=0$ slice predicted in theory \cite{Nagaosa}.


\bigskip
\section{Discussion}

\begin{figure*}
\centering
\includegraphics[width=17cm]{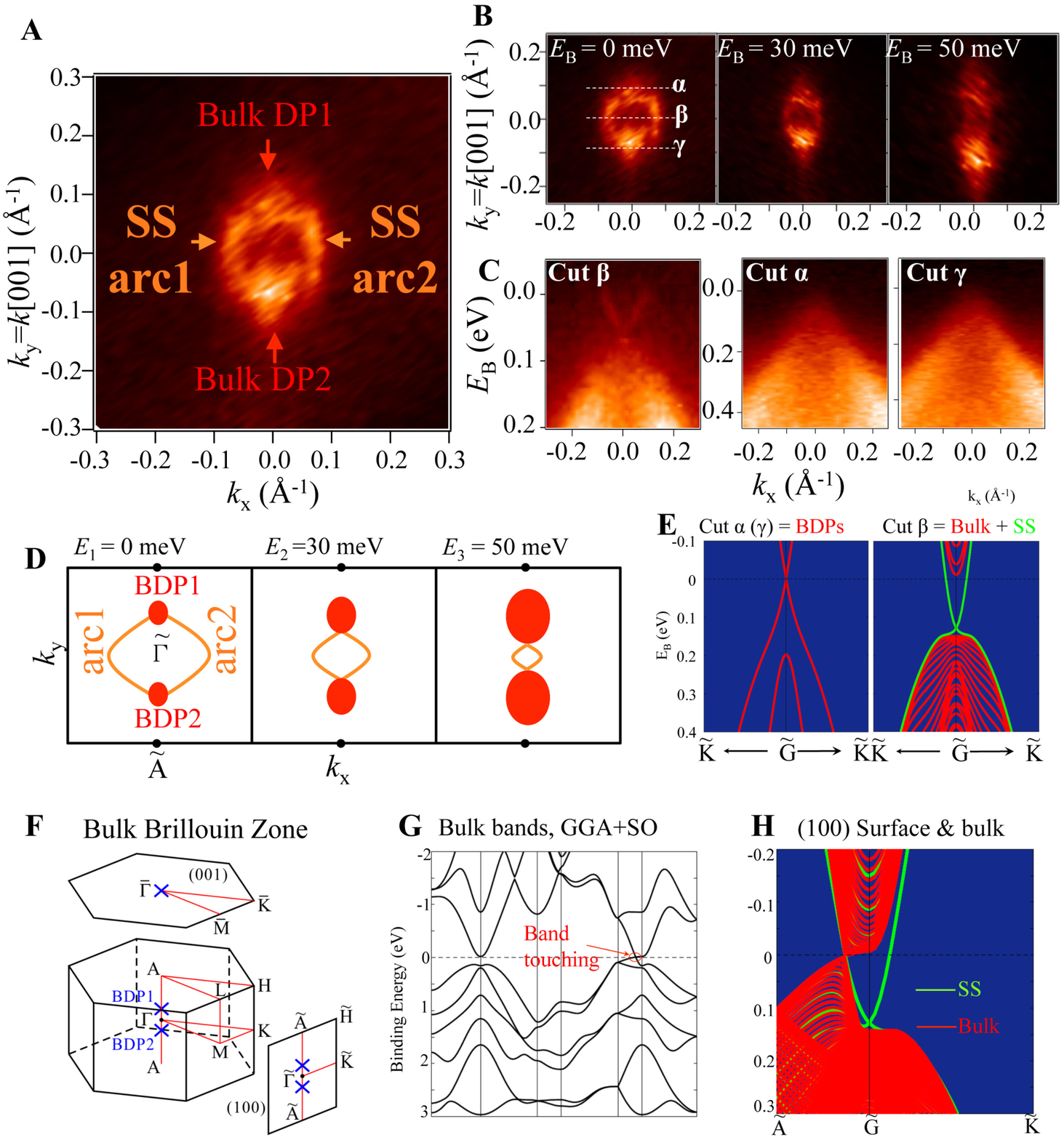}
\caption{\textbf{Observation of Fermi arc surface states.} (\textbf{A}) A Fermi surface map of the bismuth trisodium sample at photon energy of 55 eV. The BDP1 and BDP2 denote the two bulk Dirac points' momentum space locations. Two surface Fermi arcs are observed to connect these two BDPs. (\textbf{B}) APRES constant energy contours as a function of binding energy at photon energy of 55 eV. The dotted lines note the momentum space cuts for Panel (\textbf{C}). (\textbf{C}) ARPES dispersion cuts $\alpha,\beta,\gamma$ as defined in Panel (\textbf{B}) at photon energy of 55 eV. Surface states for the Fermi arcs are observed in cut $\beta$, whereas the two bulk Dirac bands are seen in cuts $\alpha,\gamma$.}\label{Arc}
\end{figure*}

The significance of the observation of Fermi arc surface states in metallic samples is analogous to the observation of Dirac surface states in 3D topological insulators \cite{RMP, Zhang_RMP} since it is the boundary modes that carry the most dramatic topological signature of a topological material \cite{Nagaosa, Ashvin2}). In fact, the topological property of Dirac semimetals was theoretically understood only very recently (see, Ref. \cite{Nagaosa, Ashvin2}), which took place after several numerical calculations and photoemission works showing that some of these materials possess bulk band structure that is analogous to 3D graphene \cite{Dirac_semi, Dai, Suyang, Oh, Chen_Na3Bi, CdAs_Hasan, CdAs_Cava}. It is theoretically shown that the most interesting or exotic physics of these materials are that of their surfaces not the bulk. Our observation of double Fermi arc surface states and their spin momentum locking demonstrates the topology of a Dirac semimetal. The observed surface states represent a new type of 2D electron gas, which is distinct from that of the surface states in a topological insulator. In a typical topological insulator such as Bi$_2$Se$_3$, the surface states' Fermi surface is a closed contour. In sharp contrast, the Fermi surface in bismuth trisodium consists of two arcs, which are bridged by the two bulk nodes (Figs.~\ref{Comparison}\textbf{E,F}). Therefore, in this novel Fermi surface topology, as one goes along the surface Fermi arc and reaches a bulk node, the wavefunction of the surface state gradually loses its surface nature and becomes a bulk band. Such exotic behavior, shown by our systematic ARPES data in Figs.~\ref{Arc}-\ref{Gapclosing}, does not exist in surface states of Bi$_2$Se$_3$ or any known topological insulators. This highlights the distinct topological nature of the nontrivial semimetal phase in a Dirac semimetal, which is fundamentally different from the known 3D TI phase \cite{RMP, Zhang_RMP}. In surface electrical transport experiments at high magnetic field, it would be interesting to study how the surface electrons wind along the arc and enter the bulk singularity \cite{Ashvin2}. Moreover, if superconductivity can be induced by bulk doping in a Dirac semimetal, the surface state superconductivity can be topologically nontrivial, which could possibly lead to double Majorana-arc states \cite{Kane_Proximity, Weyl superconductor}. Finally, the topological Dirac semimetal phase in a Dirac semimetal can be visualized as two copies of Weyl semimetals, where each bulk Dirac node is a composite of two degenerate Weyl nodes (Fig.~\ref{Comparison}\textbf{E}) that can be further split by additional symmetry breaking. There is absolutely no analogous picture for the surface states of strong topological insulators. At most, symmetry breaking in materials such as Bi$_2$Se$_3$ will gap out the surface states.

In conclusion, our systematic studies of bismuth trisodium surfaces have demonstrated a topological surface state analog for a 3D Dirac semimetal. The double surface Fermi arcs that we have observed connect the two bulk linear band touching points and exhibit spin polarization thus these results have identified the first topological phase in a Dirac semimetal, where the arc surface states and their unique spin momentum locking further evaluates a topological quantum number of $\nu_{\textrm{2D}}=1$ for the $k_{[001]}=0$ 2D (momentum slice) that uniquely defines the particular topological invariant realized in bismuth trisodium. The observed Fermi arc surface states, which represent a new type of 2D electron gas, not only open the door for studying new fundamental physics phenomena, but also offer a new class of topological materials for further engineering into the nanoscience world \cite{Ashvin_Review,Haldane, Nagaosa, 3D_Dirac, Dirac_3D, Dirac_semi, Dai, Ashvin, Ashvin2, Hosour, Ando, Ojanen, RMP, Zhang_RMP, Kim, Geim, Hasan_Na3Bi}.

\newpage
\addtocounter{figure}{-1}
\begin{figure*}[t!]
\caption{(\textbf{D}) Schematics of the constant energy contours drawn according to the theoretically calculated band structure. The red shaded areas and the orange lines represent the bulk and surface states, respectively. (\textbf{E}) Calculated band structure along Cut $\beta$ and Cut $\alpha$ ($\gamma$). (\textbf{F}) Structure of bulk and surface Brillouin zone at (001) and (100) surfaces. Bulk Dirac nodes are marked by blue crosses. Note that the two bulk Dirac cones project to the same $\bar{\Gamma}$ point on the (001) surface, while they are separated in momentum space when studied at the (100) surface. The (100) surface was not studied in previous works due to the challenge of such cleavage in ARPES experiments (\textbf{G}) First principles bulk band calculation for bismuth trisodium. (\textbf{H}) First principles calculation of the (100) surface electronic structure.}
\end{figure*}

\newpage

\begin{figure*}
\centering
\includegraphics[width=17cm]{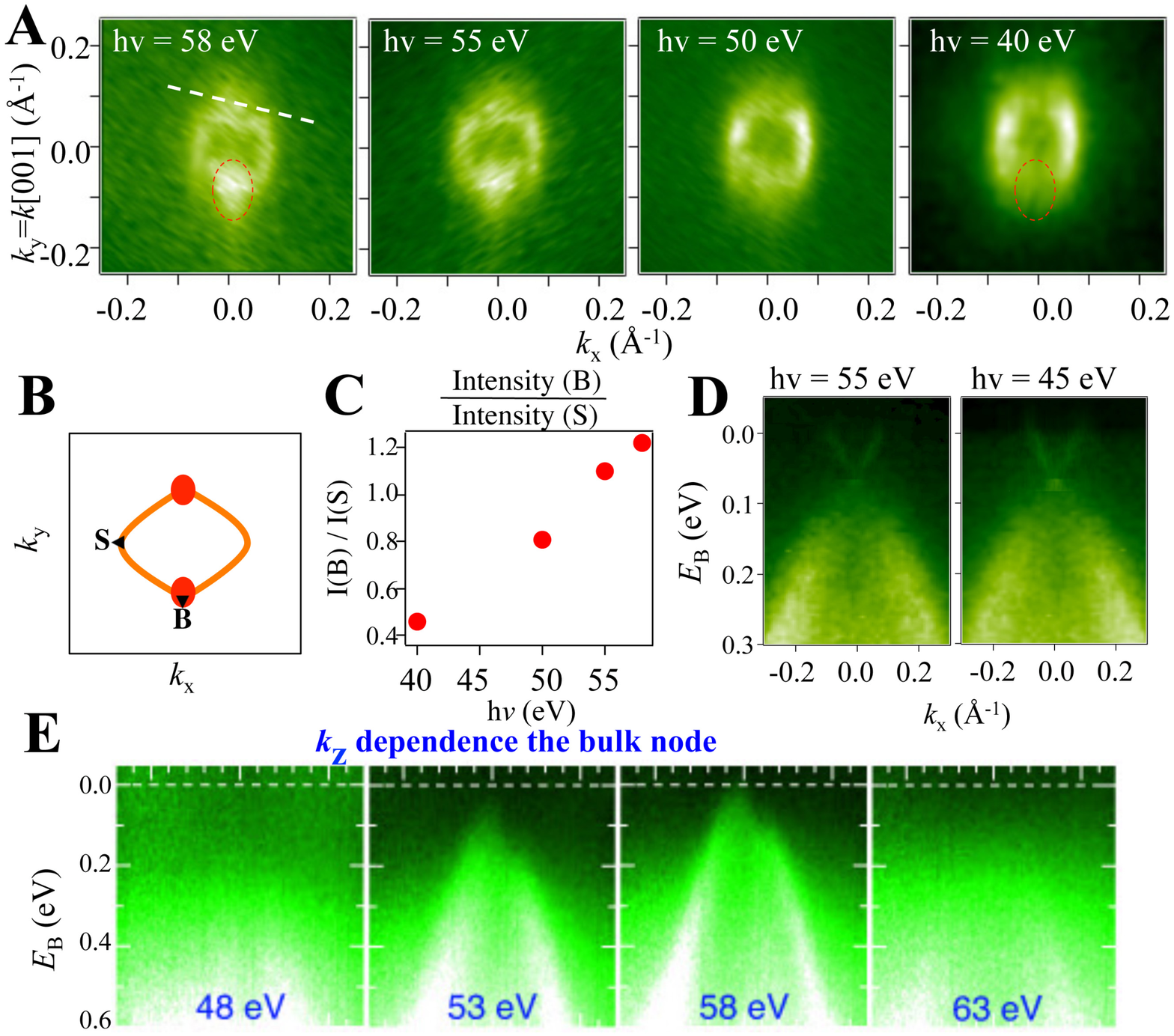}
\caption{\textbf{Systematic studies on the double Fermi arc surface states.} (\textbf{A}) ARPES Fermi surface maps at different photon energies. The double Fermi arcs are observed independent of the photon energy value, whereas the two bulk Dirac band points along $k_y$ become much weaker in intensity as the $h\nu$ is changed from 58 eV to 40 eV, revealing their surface state and bulk band nature, respectively. (\textbf{B,C}) In order to quantitatively evaluate how much the bulk band intensity gets weaker as $h\nu$ is changed from 58 eV to 40 eV (its $k_z$ dependence), we denote the middle point of the left Fermi arc as ``S'' and the bottom bulk Dirac point as ``B''. Panel (\textbf{C}) show the relative ARPES intensity between ``S'' and ``B''. The strong dependence of $\frac{I(\textrm{B})}{I(\textrm{S})}$}\label{hv_spin}
\end{figure*}
\addtocounter{figure}{-1}
\begin{figure*}[t!]
\caption{upon varying photon energy shows the strong $k_z$ dependent nature the bulk band Dirac points. (\textbf{D}) ARPES dispersion maps of the surface states at two different photon energies. The surface states with a Dirac crossing are observed at both photon energies, which further supports its 2D nature. (\textbf{E}) ARPES dispersion maps of the bulk Dirac band at different photon energies [indicated by the white dotted lines in Panel (\textbf{A})]. The bulk Dirac band only crosses the Fermi level for excitation photon energies around 58 eV, which demonstrates their three-dimensional dispersive (bulk) nature.}
\end{figure*}

\clearpage
\begin{figure*}
\centering
\includegraphics[width=17cm]{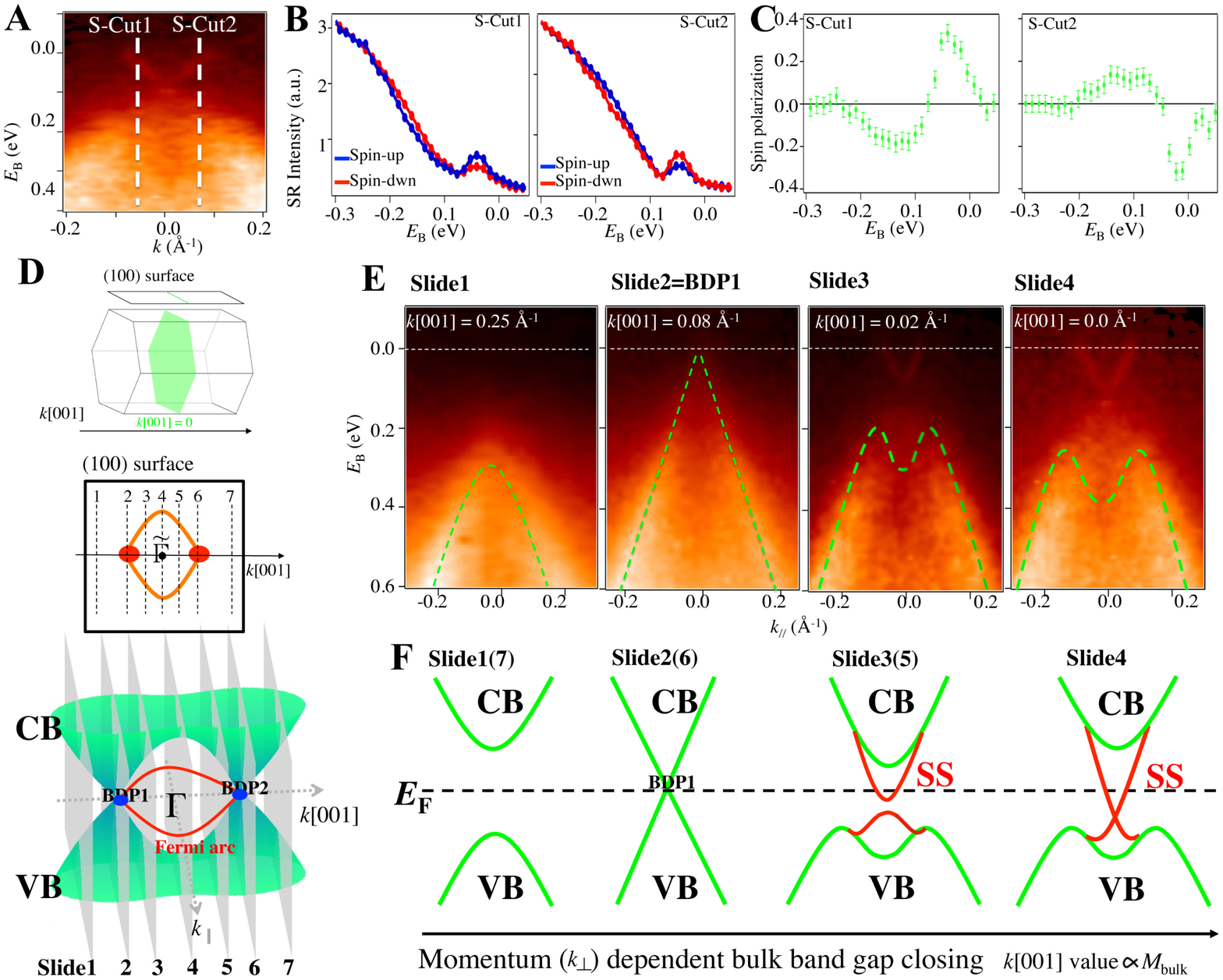}
\caption{\textbf{Surface spin polarization and Momentum-dependent band gap closing.} (\textbf{A}) The white dotted lines note the two momenta chosen for spin-resolved measurements. (\textbf{B,C}) Spin-resolved ARPES intensity and net spin polarization along the in-plane tangential direction for S-Cuts1 and 2 at photon energy of 55 eV. Clear spin polarization and spin-momentum locking are found in the data, which demonstrate the singly degenerate nature of the Fermi arc surface states along the cut $\beta$ $k$-space direction. (\textbf{D}) A schematic view of the band structure of the topological Dirac semimetal phase. Seven slices/cuts that are taken perpendicular to the $k_{[001]}$ axis are noted. (\textbf{E,F}) ARPES measured (Panel \textbf{E}) and schematic (Panel \textbf{F}) band structure for these slices are shown.}\label{Gapclosing}
\end{figure*}

\begin{figure*}
\centering
\includegraphics[width=18cm]{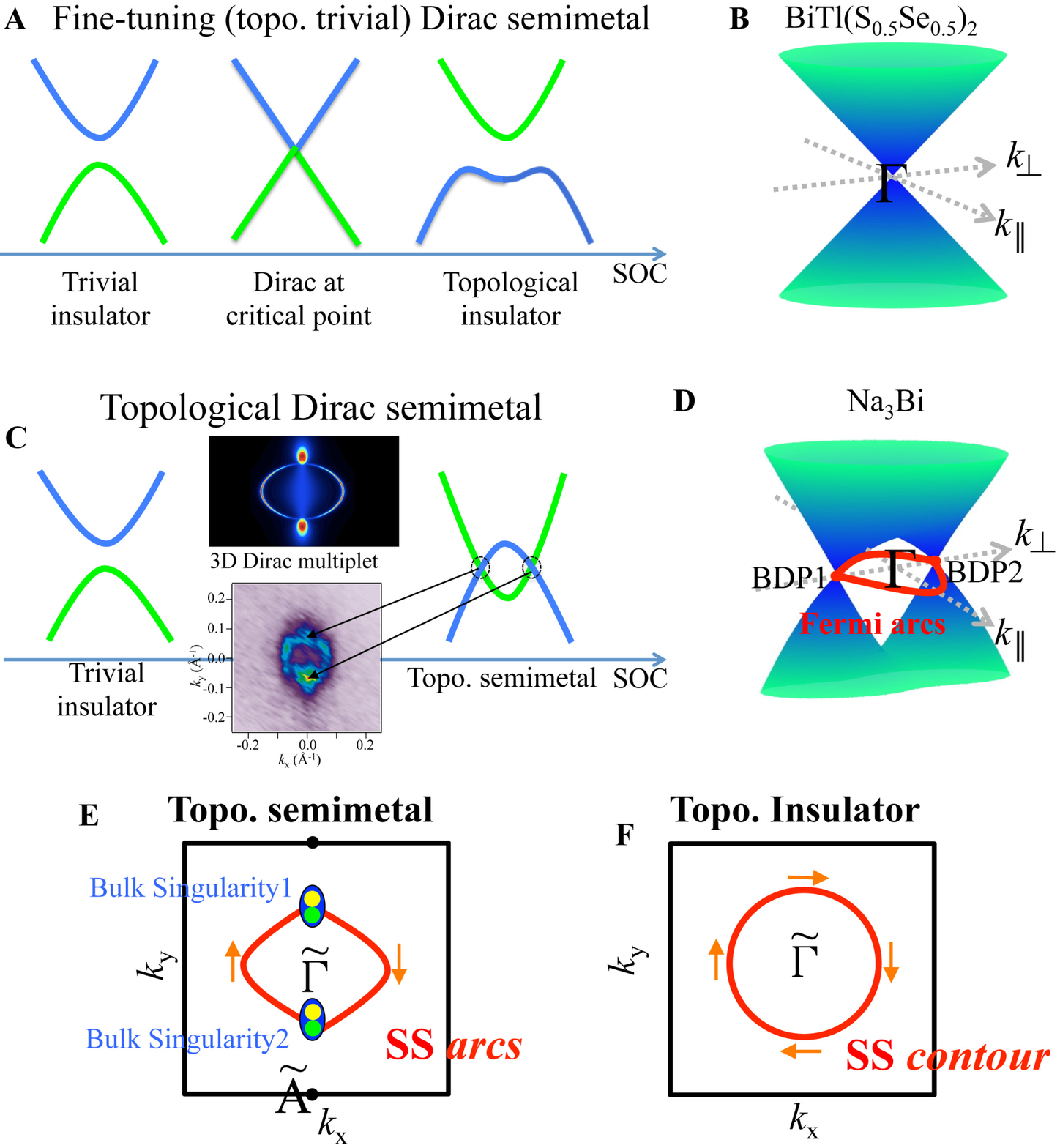}
\caption{\textbf{Comparison between the fine-tuning of a 3D Dirac semimetal and the topological Dirac semimetal.} (\textbf{A}) The 3D Dirac semimetal phase with only one bulk Dirac cone can be realized by tuning the topological phase transition to the band inversion critical point. (\textbf{B}) Cartoon band structure of the critical composition of the topological phase transition system BiTl(S$_{0.5}$Se$_{0.5}$)$_2$. (\textbf{C}) A topological Dirac semimetal phase is realized in an inverted-band structure, with additional protection by crystalline symmetries. (\textbf{D}) Schematic band structure from the}\label{Comparison}
\end{figure*}
\addtocounter{figure}{-1}
\begin{figure*}[t!]
\caption{(100) surface of a topological Dirac semimetal. The bulk band structure can be described as two bulk Dirac cones along the $k_{\perp}$ [100] direction and double surface Fermi arcs that connect the bulk Dirac nodes. In this schematic, only surface states at the bulk Dirac point energy are plotted. Surface states at other energies are not shown. (\textbf{E,F}) Schematic Fermi surface topology of bismuth trisodium and Bi$_2$Se$_3$. In bismuth trisodium, each Dirac node (the blue shaded area) can be viewed as a composite of two degenerate Weyl nodes (yellow and green areas). The orange arrows note the spin polarization according to our spin-resolved ARPES measurements. The inset of Panels (\textbf{C,E}) shows the ARPES measured Fermi surface.}
\end{figure*}

\begin{thebibliography}{99}
\bibitem{Ashvin_Review} A. M. Turner, A. Vishwanath, Beyond Band Insulators: Topology of Semi-metals and Interacting Phases. Preprint at http://arXiv:1301.0330 (2013).
\bibitem{Haldane} F. D. M. Haldane, Attachment of Surface ``Fermi Arc'' to the Bulk Fermi Surface: ``Fermi-Level Plumbing'' in Topological Metals. Preprint at http://arXiv:1401.0529 (2014).
\bibitem{Ashvin2} A. C. Potter, I. Kimchi, A. Vishwanath, Quantum Oscillations from Surface Fermi-Arcs in Weyl and Dirac Semi-Metals. Preprint at http://arXiv:1402.6342 (2014).
\bibitem{Hosour} P. Hosur, Friedel oscillations due to Fermi arcs in Weyl semimetals. \textit{Phys. Rev. B} $\mathbf{86}$, 195102 (2012).

\bibitem{Nagaosa} B.-J. Yang, N. Nagaosa, Classification of stable three-dimensional Dirac semimetals with nontrivial topology. Preprint at http://arXiv:1404.0754 (2014).
\bibitem{3D_Dirac} S. Murakami, Phase transition between the quantum spin Hall and insulator phases in 3D: emergence of a topological gapless phase. \textit{New J. Phys.} $\mathbf{9}$, 356 (2007).
\bibitem{Dirac_3D} S. M. Young \textit{et al.,} Dirac semimetal in three dimensions. \textit{Phys. Rev. Lett.} $\mathbf{108}$, 140405 (2012).
\bibitem{Dirac_semi} Z. Wang \textit{et al.,} Dirac semimetal and topological phase transitions in A$_3$Bi (A = Na, K, Rb). \textit{Phys. Rev. B} $\mathbf{85}$, 195320 (2012).
\bibitem{Dai} Z. Wang, \textit{et.al.,} Three dimensional Dirac semimetal and quantum spin Hall effect in Cd$_3$As$_2$. \textit{Phys. Rev. B} $\mathbf{88}$, 125427 (2013).

\bibitem{Ashvin} X. Wan, \textit{et al.,} Topological semimetal and Fermi-arc surface states in the electronic structure of pyrochlore iridates. \textit{Phys. Rev. B} $\mathbf{83}$, 205101 (2011).



\bibitem{Ando} M. Koshino, T. Ando, Anomalous orbital magnetism in Dirac-electron systems: Role of pseudospin paramagnetism. \textit{Phys. Rev. B} $\mathbf{81}$, 195431 (2010).
\bibitem{Ojanen} T. Ojanen, Helical Fermi arcs and surface states in time-reversal invariant Weyl semimetals. \textit{Phys. Rev. B} $\mathbf{87}$, 245112 (2013).


\bibitem{Kim} R. C. Dean, \textit{et al.,} Hofstadter's butterfly in moire superlattices: A fractal quantum Hall effect. \textit{Nature} $\mathbf{497}$, 598-602 (2013).
\bibitem{Geim} L. A. Ponomarenko, \textit{et al.,} Cloning of Dirac fermions in graphene superlattices. \textit{Nature} {\bf 497}, 594-597 (2013).
\bibitem{Hasan_Na3Bi} S.-Y. Xu \textit{et.al.,} Observation of a bulk 3D Dirac multiplet, Lifshitz transition, and nestled spin states in Na$_3$Bi. Preprint at http://arXiv:1312.7624 (2013).




\bibitem{RMP} M. Z. Hasan, C. L. Kane, Topological Insulators. \textit{Rev. Mod. Phys.} $\mathbf{82}$, 3045-3067 (2010).

\bibitem{Zhang_RMP} X.-L. Qi, S-C. Zhang, Topological insulators and superconductors. \textit{Rev. Mod. Phys.} $\mathbf{83}$, 1057-1110 (2011).
\bibitem{Wen} X.-G. Wen, Symmetry-protected topological invariants of symmetry-protected topological phases of interacting bosons and fermions.
\textit{Phys. Rev. B} $\mathbf{89}$, 035147 (2014).

\bibitem{Suyang} S.-Y. Xu, \textit{et al.,} Topological phase transition and texture inversion in a tunable topological insulator. \textit{Science} $\mathbf{332}$, 560-564 (2011).
\bibitem{Oh} M. Brahlek, \textit{et al.} Topological-Metal to Band-Insulator Transition in (Bi$_{1-x}$In$_x$)$_2$Se$_3$ Thin Films. \textit{Phys. Rev. Lett.} $\mathbf{109}$, 186403 (2012).

\bibitem{Chen_Na3Bi} Z. K. Liu \textit{et al.,} Discovery of a three-dimensional topological Dirac semimetal, Na$_3$Bi. \textit{Science} {\bf 343}, 864-867 (2014).


\bibitem{CdAs_Hasan} M. Neupane \textit{et.al.,} Observation of a topological 3D Dirac semimetal phase in high-mobility Cd$_3$As$_2$. \textit{Nature Commun.} $\mathbf{5}$, 4786 (2014).


\bibitem{CdAs_Cava} S. Borisenko \textit{et.al.,} Experimental realization of a three-dimensional Dirac semimetal. Preprint at http://arXiv:1309.7978 (2013).



\bibitem{Na3Bi_crystal} T. B. Massalski, Binary alloy phase diagrams (ASM, Materials Park, 1990).




\bibitem{Madhab} M. Neupane, \textit{et al.,} Saddle point singularity and topological phase diagram in a tunable topological crystalline insulator (TCI). Preprint at http://arXiv:1403.1560 (2014).


\bibitem{Kane_Proximity} L. Fu, C. L. Kane, Superconducting Proximity Effect and Majorana Fermions at the Surface of a Topological Insulator. \textit{Phys. Rev. Lett.} $\mathbf{100}$, 096407 (2008).
\bibitem{Weyl superconductor} T. Meng, L. Balents, Weyl superconductors. \textit{Phys. Rev. B} $\mathbf{86}$, 054504 (2012).




\end{thebibliography}
\end{document}